%% file: Template.tex
\documentclass[conference,10pt]{IEEEtran}
\IEEEoverridecommandlockouts
\usepackage{amsmath,graphicx,amsfonts}
\usepackage{booktabs}
\usepackage{multirow}
\usepackage{xcolor}
\def\BibTeX{{\rm B\kern-.05em{\sc i\kern-.025em b}\kern-.08em
    T\kern-.1667em\lower.7ex\hbox{E}\kern-.125emX}}

\begin{document}

\title{Beyond Oversmoothing: Evaluating DDPM and MSE for Scalable Speech Synthesis in ASR
\thanks{PhD funded by Huawei Technologies (UK) Co., Ltd.\newline Work supported by Google TPU Research Cloud (TRC).}
}


\author{
    \IEEEauthorblockN{
         Christoph Minixhofer, Ondrej Klejch, Peter Bell
    }
    \IEEEauthorblockA{
        \textit{Centre for Speech Technology Research, University of Edinburgh, United Kingdom} \\
        {
            \small
            \tt
            \{christoph.minixhofer,o.klejch,peter.bell\}@ed.ac.uk
        }
    }
}

%
%
%
%
\maketitle
\begin{abstract}
Synthetically generated speech has rapidly approached human levels of naturalness. However, the paradox remains that ASR systems, when trained on TTS output that is judged as natural by humans, continue to perform badly on real speech.  In this work, we explore whether this phenomenon is due to the oversmoothing behaviour of models commonly used in TTS, with a particular focus on the behaviour of TTS-for-ASR as the amount of TTS training data is scaled up. We systematically compare Denoising Diffusion Probabilistic Models (DDPM) to Mean Squared Error (MSE) based models for TTS, when used for ASR model training. 
We test the scalability of the two approaches, varying both the number hours, and the number of different speakers.  We find that for a given model size, DDPM can make better use of more data, and a more diverse set of speakers, than MSE models. We achieve the best reported ratio between real and synthetic speech WER to date (1.46), but also find that a large gap remains.

\end{abstract}
\begin{IEEEkeywords}
speech synthesis, speech recognition, evaluation, denoising diffusion
\end{IEEEkeywords}

\section{Introduction}
\label{sec:intro}

Synthetic data for model training has gained traction across several domains, including computer vision, natural language processing, and speech technology \cite{gao2023synthetic,xiang2023less,nikolenko2021synth,wood2021fake}. 
Recent works have increasingly used Text-to-Speech (TTS) systems are to produce synthetic speech for training Automatic Speech Recognition (ASR) \cite{li2018training,rosenberg2019speech,hu2022synt,casanova2022single,rossenbach2020generating,rossenbach2024problem}. 
Doing so can reduce the reliance on large volumes of real speech data, which is both difficult and expensive to obtain, and can come with ethical challenges \cite{pine2022requirements}.

As we have shown in previous work, high-quality synthetic speech—often evaluated by mean opinion scores (MOS)—does not always translate to improved ASR performance, measured by word error rate (WER) \cite{minixhofer2023evaluating}. This highlights a limitation of current TTS systems: their tendency to generate speech that sounds natural but lacks the variability of real human speech. This limitation is particularly evident in ASR training, where variability in speech patterns is crucial for effective generalization.

Previous work in TTS-for-ASR has often relied on models optimized using mean squared error (MSE) loss \cite{ren2019fastspeech,lancucki2021fastpitch}. While these models produce intelligible and stable speech \cite{pine2022requirements}, they often generate repetitive outputs that fail to capture the diversity needed for ASR. Recent research suggests that models such as Denoising Diffusion Probabilistic Models (DDPMs) \cite{ho2020ddpm}, which introduce stochasticity during generation, may better approximate the complexity of real speech by capturing a broader range of variability. By learning the reverse diffusion process, DDPMs are encouraged to generate from the entire distribution, thus producing more varied and representative synthetic speech. Nevertheless, even these models face challenges in fully bridging the gap between synthetic and real speech distributions \cite{rossenbach2024problem}.

Recent work has also explored the scaling laws of synthetic data in various domains. A plateau in test error has been identified when training large language models on synthetic data~\cite{dohmatob2024a}. Similarly, limited scaling behaviour after training on a certain number of synthetic data-points was found in computer vision~\cite{fan2024scaling}. We propose a power-law for TTS-for-ASR with respect to dataset size (Section~\ref{sec:scale}). Our findings suggest that the TTS training dataset size affects the performance of ASR models trained on synthetic data in two distinct phases. Initially, ASR performance rapidly improves as TTS learns the key aspects of speech generation. However, this is followed by a phase of diminishing returns, where additional TTS training data has minimal impact on further reducing the gap between ASR models trained on real and synthetic data.

Our work contributes two key findings: (1) We introduce a two-term power law for the relationship between dataset size used to train TTS and ASR performance when trained using synthetic data.
(2) We show that DDPM-based models show a more favorable scaling curve compared to MSE, but they encounter the same limitation of diminishing returns.

These findings indicate that while advanced models like DDPMs offer some improvements, other approaches beyond scaling synthetic data will be necessary to fully close the performance gap between synthetic and real speech data in ASR training.


\vspace{-.25em}
\section{Methodology}
\label{sec:meth}

To investigate the theoretical higher capability to capture the complex multi-modal distributions of DDPM models over MSE-based models for TTS-for-ASR, we introduce a framework which systematically evaluates both approaches.

\subsection{Modeling Text-to-Speech}

Generative models for speech synthesis can be viewed as conditional models, where the goal is to generate realistic speech conditioned on specific inputs, such as text and speaker identity \cite{huang2001analysis}. The challenge in this is ensuring that the generated speech captures the full distribution of the real speech -- one of the failure modes when aiming to generate diverse outputs is mode collapse, where the model fails to generate the full range of possible outputs and instead focuses on a limited subset. In MSE-based models, mode collapse manifests as oversmoothed, low-variance outputs that do not capture the full variability of real speech. In contrast, the stochastic nature of DDPMs, combined with their ability to model complex distributions, should inherently mitigate the risk of mode collapse.

\subsection{Mean Squared Error (MSE) Models}
\label{sec:mse}

Let \(\mathbf{x}\) denote the target speech feature (e.g., a mel spectrogram) and \(\mathbf{c}\) represent the conditioning variables, including textual input (as a phone sequence) and speaker identity (as a vector). The goal of the TTS model $f_\theta$ is to learn the conditional distribution \(p(\mathbf{x} \mid \mathbf{c})\). For MSE-based models, this is typically achieved by minimizing the L2 loss between the generated output \(f_\theta(c)\) and the ground truth \(\mathbf{x}\):

\begin{equation}
L_{\text{MSE}}(\theta) = \mathbb{E}_{\mathbf{x}, \mathbf{c}} \left[\|\mathbf{x} - f_\theta(c)\|_2^2\right]
\end{equation}

This approach assumes that the conditional distribution \(p(\mathbf{x} \mid \mathbf{c})\) can be adequately captured by a single mode, which may lead to oversmoothing, particularly in cases where the true distribution is multimodal. This is often referred to as the one-to-many problem in TTS \cite{ren2019fastspeech}.

\subsection{Denoising Diffusion Probabilistic Models (DDPMs)}
\label{sec:ddpm}

In contrast, DDPMs seek to model the entire distribution \(p(\mathbf{x} \mid \mathbf{c})\) by introducing a stochastic process. First, a \textit{forward process} progressively adds Gaussian noise to the data over a series of steps, degrading the data into a noise distribution. Then, a reverse process is learned, which gradually removes the noise at each step, reconstructing the original data from the noisy version. This approach differs fundamentally from traditional Mean Squared Error (MSE)-based models used in speech synthesis by focusing on learning this noise-reversal mechanism instead of directly predicting waveforms.



In the forward diffusion process, Gaussian noise is gradually added to the target speech feature \(x_0\), such as a mel spectrogram, over a series of timesteps \(n=1,2,\dots,N\). The result is a progressively noisier version of the speech feature, denoted as \(x_n\), which approaches a Gaussian distribution as \(n\) reaches the maximum value \(N\). This forward process is defined as:
\begin{equation}
q(\mathbf{x}_n \mid \mathbf{x}_{n-1}) = \mathcal{N}(\mathbf{x}_n; \sqrt{1 - \beta_n} \mathbf{x}_{n-1}, \beta_n \mathbf{I})
\end{equation}
where 
\(\beta_n\) represents the variance schedule for noise at each timestep of the forward diffusion process.

To generate new data, DDPMs learn the reverse diffusion process, which denoises \(x_N\) step-by-step back to \(x_0\). The reverse process is parameterized by a neural network \(p_\theta(\mathbf{x}_{n-1} \mid \mathbf{x}_n)\), which predicts the mean $\mu_\theta(\mathbf{x}_n, n)$ and variance $\Sigma_\theta(\mathbf{x}_n, n)$ of a Gaussian distribution based on current noised speech features $\mathbf{x}_n$ and timestep $n$:
\begin{equation}
p_\theta(\mathbf{x}_{n-1} \mid \mathbf{x}_n) = \mathcal{N}(\mathbf{x}_{n-1}; \mu_\theta(\mathbf{x}_n, n), \Sigma_\theta(\mathbf{x}_n, n))
\end{equation}

\subsubsection*{Conditioning} In conditional DDPMs, the model learns to generate speech by conditioning the diffusion process on specific inputs such as text (as a phone sequence) and speaker identity (as a d-vector \cite{wan2018generalized}). Let \(\mathbf{c}\) represent the conditioning variables, including both the text and speaker information. The conditional reverse process is then modeled as:
\begin{equation}
p_\theta(\mathbf{x}_{n-1} \mid \mathbf{x}_n, \mathbf{c}) = \mathcal{N}(\mathbf{x}_{n-1}; \mu_\theta(\mathbf{x}_n, n, \mathbf{c}), \Sigma_\theta(\mathbf{x}_n, n, \mathbf{c}))
\end{equation}
At each timestep \(n\), the model takes both the noised input \(x_n\) and the conditioning information $c$ to predict the parameters of the distribution from which the next denoised sample \(x_{n-1}\) is drawn. By conditioning the generation on text and speaker identity, the model ensures that the synthetic speech is not only realistic but also aligned with the intended content (text) and voice characteristics (speaker).

During training, DDPMs optimize the following objective to predict the noise \(\epsilon_n\) added in each step, conditioned on \(c\):

\begin{equation}
L_{\text{DDPM}}(\theta) = \mathbb{E}_{\mathbf{x}_0, \epsilon, n, \mathbf{c}} \left[\|\epsilon - \epsilon_\theta(\mathbf{x}_n, n, \mathbf{c})\|_2^2\right]
\end{equation}
In this formulation,  \(\epsilon_\theta(\mathbf{x}_n, n, \mathbf{c})\) is the model’s prediction of the noise added at timestep \(n\), given the conditioning information \(c\).

\subsection{Model Architectures}

\begin{figure}[t]
     \centering
     \includegraphics[width=1\linewidth]{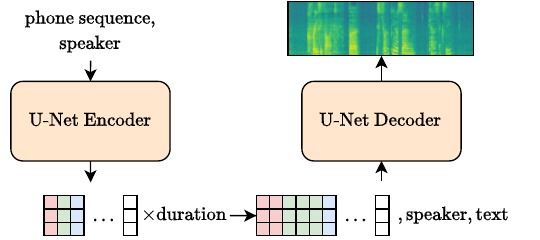}
     \caption{The architecture of the TTS system. The encoder outputs a prosody representation on phone level, while the decoder outputs a frame-level mel spectrogram.}
     \label{fig:architecture}
     \vspace{-1em}
 \end{figure}


We used an architecture consisting of two U-Net \cite{ronneberger2015unet} models: one for generating prosodic features and another for producing mel spectrograms. Both are trained using the MSE and diffusion approaches outlined in Sections \ref{sec:mse} and \ref{sec:ddpm}.

\subsubsection*{U-Net for Prosody Representation}

The first U-Net, denoted as \( \text{U-Net}_{\text{ENC}} \), generates a two-dimensional representation of prosody, utilizing the Continuous Wavelet Transform (CWT) of pitch, energy as in FastSpeech \cite{ren2019fastspeech}. To arrive at a unified target for generation, we also include the CWT of the phone duration. The prosody representation \( \mathbf{P} \) is conditioned on the phone sequence and speaker identity:

\begin{equation}
\mathbf{P} = \text{U-Net}_{\text{ENC}}(\text{phone sequence}, \mathbf{s})
\end{equation}

where \(\mathbf{s}\) is a d-vector representing the speaker. The prosody features capture both temporal and frequency-domain characteristics, providing a multi-resolution analysis that is crucial for the subsequent synthesis stages.

\subsubsection*{U-Net for Mel Spectrogram Generation}

The second U-Net, \( \text{U-Net}_{\text{DEC}} \), transforms the prosody representation into a mel spectrogram, which serves as the final output. It is additionally conditioned on semantic features derived from the text using a pre-trained Flan-T5-Base \cite{chung2024scaling} model.

\begin{equation}
\mathbf{M} = \text{U-Net}_{\text{DEC}}(\mathbf{P}', \mathbf{s},\text{text features})
\end{equation}

where \( \mathbf{P}' \) is the expanded prosody matrix (repeated along the time axis according to the predicted durations) and \( \mathbf{M} \) is the generated mel spectrogram.

\subsection{Scaling Laws for TTS-for-ASR}
\label{sec:scale}
We introduce a scaling law framework to describe how TTS-for-ASR performance in \(\text{WERR}\) scales with the dataset size \(D\). As training data increases, the model performance could theoretically reach, but not surpass, 1, since this would be equivalent to training the ASR model with real data \cite{minixhofer2023evaluating}.

We hypothesise that the performance follows a two-regime scaling law with respect to dataset size. This scaling law consists of the \textit{variance-limited} phase parametrized by $\alpha$ in which any additional data used to train TTS improves ASR as the new data it allows model to better approximate the target speech distribution. This is then folowed by a \textit{resolution-limited} regime parametrized by $\gamma$ in which the TTS models' complexity limits any further increases in usefulness of the synthetic speech for ASR.

\begin{equation}
\text{WERR}(D)\propto D^{-\alpha}+D^{-\gamma}
\end{equation}

Simpler methods may exhibit higher \(\alpha\) leading to rapid improvements with limited data. More complex and stochastic methods may lead to higher \(\gamma\), requiring more data but delaying the slowdown in performance gains. We can aditionally fit a function \(S\) with scalars \(A\) and \(B\) to the results of a model with paramters \(\theta\) to derive approximate values for \(\alpha\) and \(\gamma\). We can use this function to approximately predict the scaling behaviour when increasing dataset size further.
\begin{equation}
S_\theta(D)=AD^{-\alpha}+BD^{-\gamma}
\end{equation}

\subsection{Evaluation}

In evaluation we focus on the models' ability to generate synthetic speech that is suitable for ASR training, particularly as the amount of training data and speaker diversity increases. We utilize the WER ratio (WERR) as the primary metric, defined as the WER of an ASR system trained on synthetic data (WER-Synth) divided by the WER of the same system trained on real data (WER-Real). This ratio provides a direct measure of how well the synthetic speech matches the characteristics of real speech in terms of ASR performance \cite{minixhofer2023evaluating}.

\vspace{-.25em}
\section{Experiments}
\label{sec:exp}

We set out to evaluate the effectiveness of Denoising Diffusion Probabilistic Models (DDPM) compared to Mean Squared Error (MSE) models in generating synthetic speech for Automatic Speech Recognition (ASR) training. Specifically, we aim to assess how these models perform when scaling the amount of training data and increasing speaker diversity. We hypothesize that DDPM models, due to their probabilistic nature, will outperform MSE models in scenarios with larger datasets and greater speaker diversity.

\subsection{Experimental Setup}

\subsubsection*{Datasets}
All datasets used in these experiments are derived from the LibriHeavy \cite{kang2024libriheavy} corpus. To ensure fair comparisons, we create three subsets for each dataset size and speaker diversity condition: (1) a subset for training the TTS system, (2) a subset for training the ASR system, and (3) a subset for evaluating the ASR system. Within each experiment, the same set of speakers and their proportions are maintained across all three subsets, ensuring consistency in speaker representation. However, the specific speakers and their proportions vary between different experiments. This variability across experiments is acceptable because we focus on comparing the word error rate (WER) ratio, which measures the relative performance of the ASR system, rather than absolute WER. Additionally, there is no overlap in transcripts between the TTS training, ASR training, and ASR evaluation sets, ensuring that performance differences are attributed to variations in the TTS models and dataset conditions, rather than differences in the evaluation data.

\subsubsection*{TTS Model Configuration}  
The TTS systems used in these experiments are based on a U-Net architecture, with configurations tailored for both DDPM and MSE models. Both models also incorporate a Flan-T5-Base \cite{chung2024scaling} language model for conditioning on the original text using cross-attention.

The \textbf{MSE} model minimizes the mean squared error between the generated and true mel spectrograms (see Section~\ref{sec:mse}).

The \textbf{DDPM} model is trained with a diffusion loss function, learning to predict the noise added during the diffusion process (see Section~\ref{sec:ddpm}).

All models were trained using a batch size of 16 and used a cosine learning rate schedule starting at $4 \times 10^{-5}$. The training was conducted over 500,000 iterations with an exponential moving average (EMA) decay rate of 0.9999 to stabilize the model parameters. For inference, a DDIM sampler with 20 sampling steps was used, applying a classifier-free guidance weight of 7.5 and a rescale factor of 0.7 to control over-exposure during generation. In line with previous work, the training utilized a rescaled noise schedule to ensure zero terminal signal-to-noise ratio (SNR), and the inference was performed starting from the last timestep, ensuring congruency between the training and inference processes \cite{lin2024common}. We make our code and detailed configuration publicly available.

\vspace{-.25em}
\section{Results \& Discussion}

\begin{figure}[t]
    \centering
    \includegraphics[width=\linewidth]{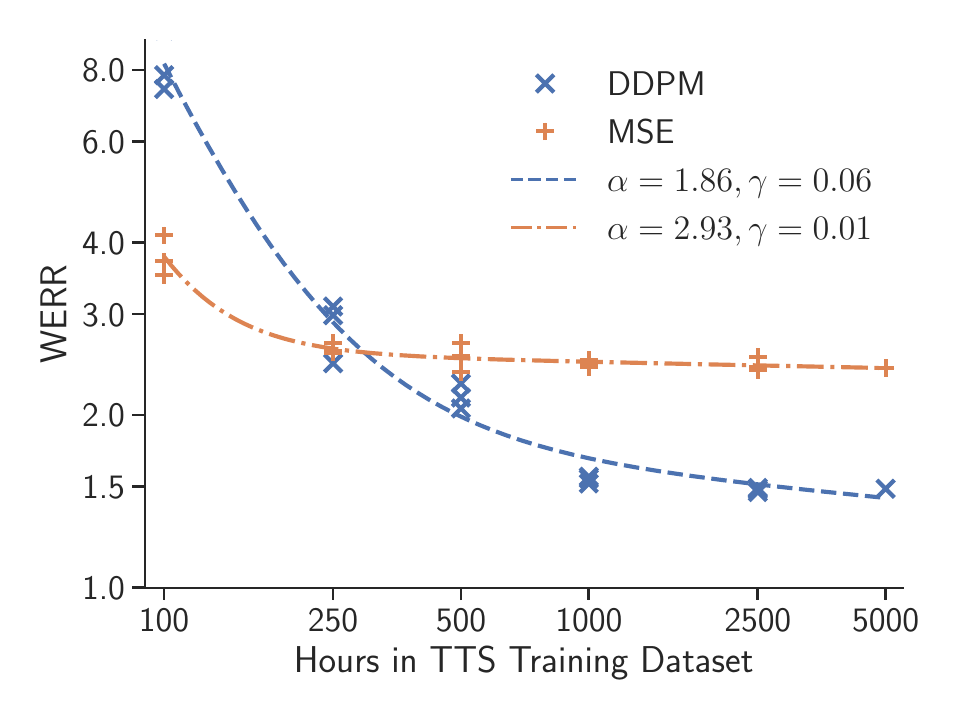}
    \caption{The performance of ASR systems trained on DDPM and MSE TTS models with different amounts of training data, and the underlying power laws (fit using a scalar for each term).}
    \label{fig:ddpmvsmse}
    \vspace{-1em}
\end{figure}

Here, we present the performance results of two models—Mean Squared Error (MSE) and Denoising Diffusion Probabilistic Models (DDPM)—across varying dataset sizes and levels of speaker diversity. We particularly focus on the models' scalability with data and speaker diversity, in terms of the Word Error Rate (WER) ratio achieved.

\subsection{Data Scaling and Performance Differences}

As shown in Figure~\ref{fig:ddpmvsmse}, the MSE and DDPM model exhibited vastly differing scaling properties. The MSE model, while performing reasonably well with smaller datasets, showed limited improvement as the dataset size increased. Even when scaled to much larger datasets, the MSE model did not approach the desired performance threshold of WER ratios close to 1. This suggests that, for the MSE model, further scaling of data offers diminishing returns, rendering it less suitable for scenarios requiring vast amounts of training data. We hypothesize that this limitation stems from the model's intrinsic bias toward more oversmoothed outputs, which fail to leverage the additional variance present in larger datasets.

In contrast, the DDPM model showed different scaling behaviour. While initially underperforming in smaller data regimes (\(\approx300\) hours and less), the model showed significant improvements as the dataset size increased. 

\subsection{Speaker Diversity}
Next, we investigate the relationship between speaker diversity and model performance. Across all dataset sizes, MSE models consistently outperformed DDPM models when trained on smaller datasets (\(\approx100\) hours), regardless of the level of speaker diversity. For example, at 100 hours, the MSE model achieved WER ratios of 3.62, 3.66, and 3.92 for low, medium, and high diversity, respectively. By comparison, the DDPM model produced WER ratios of 8.07, 8.33, and 7.44 under the same conditions. These results suggest that, in lower data regimes, the more deterministic nature of the MSE model is advantageous.
However, this trend shifted as dataset sizes increased. With 500 hours of training data and beyond, the highest speaker diversity always performed best for the DDPM model. However, as can be seen in Table~\ref{tab:splits}, the relative difference between the lowest and highest diversity diminished as training data size increased, from 8\% with 100 hours to 4\% for 5000 hours. This indicates that a similar effect to the diminishing returns in dataset size also applies to speaker diversity.

\subsection{The Scaling Function}

\input{figures/results_table}

We additionally investigate the values of \(\alpha\) and \(\gamma\) when fit to the model results as described in Section~\ref{sec:scale}. The better scaling properties in the inital, variance-limited stage of the MSE model are reflected in a higher \(\alpha\) (2.93) than for the DDPM model (1.86). However, \(\gamma\) was significantly lower for the MSE model (0.01), leading to earlier stagnation in WERR than for the DDPM model (0.06).
However, even with its improved scaling properties, and our optimistic assumption that the model never completely stagnates and keeps following the power-law, at least a million hours of synthetic training data would be required to reach the same performance as using real data. Further research is needed to explain this gap.

\vspace{-.25em}
\section{Conclusion}

Our findings demonstrate that while MSE models perform well with smaller datasets, their scalability is limited, resulting in diminishing performance improvements as data increases. In contrast, DDPM models exhibit better scalability, particularly with large and diverse datasets, though the data requirements to achieve real-speech performance remain significant. Overall, DDPM models provide more potential for future large-scale speech synthesis applications, particularly as larger datasets become more accessible. However, our proposed scaling law indicates that they would require at least one million hours of synthetic speech to match the performance of real speech, far exceeding the size of available open datasets \cite{kang2024libriheavy,pratap2020mls,chen2021gigaspeech}. Additionally, while DDPM models handle speaker diversity better with larger datasets, the diminishing returns observed in both data size and diversity highlight the need for alternative approaches to further reduce the performance gap between synthetic and real speech in ASR tasks. Future work should explore methods beyond dataset scaling to overcome these limitations.

\bibliographystyle{IEEEbib}
\bibliography{Template}

\end{document}

%% file: figures/results_table.tex
\begin{table}[]
\begin{tabular}{@{}cccccccc@{}}
\toprule
\multirow{2}{*}{\textbf{Hrs.}} & \multirow{2}{*}{\textbf{Div.}} & \multirow{2}{*}{\textbf{Spk.}} & \multicolumn{2}{c}{\textbf{WER-Ratio ↓}}           & \multicolumn{2}{c}{\textbf{WER ↓}}                 & \multirow{2}{*}{\textbf{\begin{tabular}[c]{@{}c@{}}Real\\ WER ↓\end{tabular}}} \\ \cmidrule(lr){4-7}
                                  &                                     &                                    & \textbf{DDPM} & \multicolumn{1}{c|}{\textbf{MSE}}  & \textbf{DDPM} & \multicolumn{1}{l|}{\textbf{MSE}}  &                                                                                \\ \midrule
\multirow{3}{*}{100}             & Low                                 & \multicolumn{1}{c|}{25}            & 8.07          & \multicolumn{1}{c|}{\textbf{3.62}} & 78.4          & \multicolumn{1}{c|}{\textbf{35.2}} & 9.72                                                                           \\
                                  & Medium                              & \multicolumn{1}{c|}{40}            & 8.33          & \multicolumn{1}{c|}{3.66}          & 80.3          & \multicolumn{1}{c|}{35.3}          & 9.64                                                                           \\
                                  & High                                & \multicolumn{1}{c|}{62}            & 7.44          & \multicolumn{1}{c|}{3.92}          & 71.3          & \multicolumn{1}{c|}{37.6}          & 9.59                                                                           \\ \midrule
\multirow{3}{*}{250}             & Low                                 & \multicolumn{1}{c|}{60}            & 3.09          & \multicolumn{1}{c|}{2.58}          & 30.2          & \multicolumn{1}{c|}{25.2}          & 9.76                                                                           \\
                                  & Medium                              & \multicolumn{1}{c|}{100}           & 2.98          & \multicolumn{1}{c|}{2.66}          & 29.0          & \multicolumn{1}{c|}{25.9}          & 9.73                                                                           \\
                                  & High                                & \multicolumn{1}{c|}{150}           & 2.46          & \multicolumn{1}{c|}{2.56}          & \textbf{23.7} & \multicolumn{1}{c|}{24.7}          & 9.64                                                                           \\ \midrule
\multirow{3}{*}{500}             & Low                                 & \multicolumn{1}{c|}{125}           & 2.29          & \multicolumn{1}{c|}{2.39}          & 22.4          & \multicolumn{1}{c|}{23.4}          & 9.79                                                                           \\
                                  & Medium                              & \multicolumn{1}{c|}{200}           & 2.19          & \multicolumn{1}{c|}{2.58}          & 21.1          & \multicolumn{1}{c|}{24.9}          & 9.65                                                                           \\
                                  & High                                & \multicolumn{1}{c|}{310}           & \textbf{2.04} & \multicolumn{1}{c|}{2.65}          & \textbf{19.7} & \multicolumn{1}{c|}{25.5}          & 9.62                                                                           \\ \midrule
\multirow{3}{*}{1000}            & Low                                 & \multicolumn{1}{c|}{250}           & 1.58          & \multicolumn{1}{c|}{2.45}          & 15.5          & \multicolumn{1}{c|}{24.0}          & 9.79                                                                           \\
                                  & Medium                              & \multicolumn{1}{c|}{400}           & 1.58          & \multicolumn{1}{c|}{2.55}          & 15.4          & \multicolumn{1}{c|}{24.9}          & 9.76                                                                           \\
                                  & High                                & \multicolumn{1}{c|}{620}           & \textbf{1.53} & \multicolumn{1}{c|}{2.48}          & \textbf{14.8} & \multicolumn{1}{c|}{24.1}          & 9.71                                                                           \\ \midrule
\multirow{3}{*}{2500}            & Low                                 & \multicolumn{1}{c|}{625}           & 1.52          & \multicolumn{1}{c|}{2.42}          & 15.0          & \multicolumn{1}{c|}{23.9}          & 9.87                                                                           \\
                                  & Medium                              & \multicolumn{1}{c|}{956}           & 1.47          & \multicolumn{1}{c|}{2.52}          & 14.4          & \multicolumn{1}{c|}{24.8}          & 9.84                                                                           \\
                                  & High                                & \multicolumn{1}{c|}{1531}          & \textbf{1.46} & \multicolumn{1}{c|}{2.59}          & \textbf{14.2} & \multicolumn{1}{c|}{25.2}          & 9.72                                                                           \\
5000                             & Low                                 & 1531                               & 1.50          & 2.42                               & \textbf{15.0} & 24.2                               & 10.0                                                                           \\ \bottomrule
\end{tabular}
\caption{Results for different dataset sizes and speaker diversities.}
\label{tab:splits}
\vspace{-.5em}
\end{table}